\documentstyle[12pt]{article}

\begin{document}

\thispagestyle{empty}  

\begin{flushright}   
FTUV-IFIC 99-11                      
\end{flushright}   

\begin{center}
{\Large{\bf Operator approach to the Gluing\\
\vspace{0.3cm}
Theorem in String Field Theory}}
\end{center}

\vspace{1cm}
\begin{center}

{\large Abdulmajeed ABDURRAHMAN}\\
abdurrahman\,@\,v2.rl.ac.uk \\
{\it Rutherford Appleton Laboratory,\\
  Chilton, Didcot, Oxon OX11 0QX, United Kingdom}\\
\vspace{.2cm}
{\large Jos\'e BORDES\footnote{Also at IFIC, Universitat de Valencia-CSIC
(Centro Mixto)}}\\
jose.m.bordes\,@\,uv.es\\
{\it Dept. Fisica Teorica, Univ. de Valencia,\\
  c. Dr. Moliner 50, E-46100 Burjassot (Valencia), Spain}\\
and \\
{\large Crist\'obal LARA}\\
{\it Dept. Fisica Teorica, Univ. de Valencia,\\
  c. Dr. Moliner 50, E-46100 Burjassot (Valencia), Spain}\\
\end{center}

\vspace{3cm}

\begin{abstract}
{\small{
An algebraic proof of the Gluing Theorem at tree level of perturbation
theory in String Field Theory is given. Some
applications of the theorem to  closed string non-polynomial action
are briefly discussed}}
\end{abstract}

\vspace{1cm}

\noindent
Pacs: 11.10.-z,11.25.Sq

\newpage
The gluing of vertices in String Field Theory and in particular in the
case of closed strings, is at the basis of the construction of a 
perturbative approach to the theory.  The proof of various properties
such as associativity,  gauge invariance of the action, etc., makes use
 of such gluing of vertices. To tackle the problem, Peskin,
Le Clair and Preitchschopf \cite{ALE} have developed and proved a 
theorem which shows how to construct a given vertex starting from  an
elementary one. In principle, one could start from the simplest 
vertex with three strings and generate vertices for any number of strings. With
some precisions about the integration over the modular space it will
give the starting point to build a field theory for closed strings with
only one term, namely, the three strings vertex which appropriately
will generate the infinite non-polynomial action \cite{MKA}.

In the proof of the theorem in \cite{ALE}, the authors use the Riemann surfaces
with holes representation of vertices to show that the contraction
of legs in different vertices (which can be represented by the BPZ 
inner product \cite{ABE}) gives another vertex representation the
Riemann surface arising from sewing two surfaces around the hole
boundaries.

Moreover, it has been suggested in \cite{CHA,NONPOL} that the scattering
amplitude of N strings, at any order in perturbation theory can be interpreted as
the gluing of elementary vertices. In this approach to the field
theory proposed by Witten \cite{EWI}, the propagator between
vertices is simulated by the joining of  intermediate strings in all
possible parameterizations. This requires the formulation of the theory
in terms of the left and right hand pieces of the string.

Following this suggestion we prove the gluing theorem using the operator
approach (to string vertices) developed in \cite{JBC}, in which strings
are expanded using left and right degrees of freedom and the Neumann
functions are written in terms of infinite dimensional matrices.
These matrices are just the ones appearing in the canonical transformation
from the usual self-energy Fock space operators to the ones splitting
left and right degrees of freedoms as mentioned above.

Using general properties of these matrices, the proof of the theorem
simplifies. One can then explicitly generate the Neumann function
of the N-strings vertex starting from the 3-strings vertex appearing
in the open String Field Action. In closed strings, in order to get the
non-polynomial expansion, one needs to know how to contract the
restricted vertex interaction \cite{MKA}. Following \cite{CHA} where the
closed string interaction is interpreted as the reparametrization of the
string, it appears to be possible \cite{AAB} to perform a generalization of the
gluing of vertices (which gives just a point in the modular space)
to the gluing of parametrized vertices covering a region of the modular
space.

The  organization of this note is as follows. First we review
the form of the 
vertex when the midpoint string interaction is explicitly
considered. Then we briefly state the Gluing Theorem of \cite{ALE}.
The proof of the theorem for arbitrary vertices is sketched, and the
relevant steps, in the case of the gluing of two vertices with
3 strings to give  the one with four (3 + 3 $\rightarrow$ 4),
are discussed in some detail. We will see that in the critical
dimension, the ghost degrees of freedom cancel the determinant of the
Laplacian coming from the gluing procedure, represented here as the
determinant of an infinite dimensional matrix. Finally we state our
conclusions and future outlook. 

When string fields are written in terms of half string oscillator modes,
the N-string vertex takes on a very elegant and simple form. In the spirit
of the original work of \cite{EWI} in String Field Theory, the
interaction vertex is obtained by identifying oscillators 
referring to opposite halves of two adjacent strings. It has been
shown in \cite{JBA} that the Fourier components of the Neumann functions
for the N-strings vertex can be written in terms of
particular linear combinations of the two infinite dimensional matrices
relating the half-string degrees of freedom to the conventional Hamiltonian
eigenstates oscillator modes. For the purpose of this work, this
particular algebraic representation of the Neumann functions is very
useful; the reason is, in order to prove the theorem one needs to 
calculate explicitly the inverse of some combinations of these
matrices which have been calculated before \cite{JBA} in
its most general form.

Writing the N-strings vertex in an operator form, and splitting the
vertex into coordinate and ghost sector, we get for the coordinates 
degrees of freedom

\begin{equation}
< \,  V_x (1 \cdots N)| = \Pi^N_{i=1}
\delta \left(\sum_{i = 1}^N k_i\right) \,  _i< k_i | V_x (1 \cdots
 N).
\end{equation}

The $N$ momentum states $_i < k_i|$ are eigenstates of the 
corresponding i-string 
momentum operator $a^{(i)}_0$ and the vertex operator, written in
terms of string creation and annihilation
oscillator modes, is given by

\begin{equation}
V_x (1 \cdots N) = exp \left\{ \frac{1}{2} \sum_{r,s} \sum_{n,m}
a_{\mu,n}^{(r)} \, N_{n,m}^{rs}
\, a_m^{\mu (s)} \right\}.
\label{xvertex}
\end{equation}
\noindent
In $D$ space-time dimensions $
\mu = 1, \cdots D$ and  $r,s = 1, \cdots, N$.
$N_{n, m}^{rs}$ are the Fourier components of the relevant Neumann
function.

For the ghost degrees of freedom we use the bosonic formalism. Hence,
one has  for the ghost vertex a similar form to the matter vertex

\begin{equation}
< V_{\phi} (1 \cdots N)| = \, \delta \left(\sum_{i = 0} q_i + Q\right)
\Pi_{i=1}^N
 \, \,
_i\;< - q_i - Q| \, V_\phi (1 \cdots N).
\end{equation}
\noindent
$Q$ is the weight of the curvature  term in the string action 
\cite{MBG} and the states $ _i< -q_i - Q|$ are ghost number eigenstates. 
In terms of the oscillator modes ($\phi_n^{(r)}$) for the bosonic
reparametrization ghosts one has 

$$
< - q - Q| \phi_0 = < - q - Q| q 
$$
The vertex operator now takes the form

\begin{equation}
\begin{array}{ll}
V_\phi (1, \cdots,  N)  = & exp  \left\{ \frac{1}{2} 
\sum_{r \neq s} \phi_0^{(r)} \, \tilde{N}_{0,0}^{r,s} 
\phi_0^{(s)} +  \right. \\[2ex]
&  \left. + \sum_{r,s} \sum_n \phi_0^{(r)} \tilde{N}_{0,n}^{r,s} 
\phi_n^{(s)} + 
 \frac{1}{2} \sum_{r,s} \sum_{n,m} \phi_0^{(r)}
\tilde{N}^{r,s}_{n,m} \phi_m^{(s)} \right\}.
\end{array}
\end{equation}

Notice that the ghost vertex differs from (\ref{xvertex}) in the ghost
number quadratic term as well as by having an extra linear term.
Coordinate and ghost coefficients are related according to

\begin{equation}
\begin{array}{l}
\tilde{N}_{0,0}^{r,s}  = N^{r,s}_{0,0} - \frac{1}{2} N_{0,0}^{r,r}
- \frac{1}{2} N_{0,0}^{s,s}, \\[2ex]
\tilde{N}_{0,n}^{r,s} = N_{0,n}^{r,s} - \frac{1}{2} K_n^s,\\[2ex]
\tilde{N}_{n,m}^{r,s} = N_{n,m}^{r,s},
\end{array}
\end{equation}

\noindent
where the linear coefficient represented by the vector $\vec{K}^s$ 
has as components:
\begin{equation}
K_n^s = -
\frac{4}{3} \frac{(-1)^{\frac{n}{2}}}{(2 n)^{\frac{1}2}} (1 + (-1)^n).
\label{vector}
\end{equation}

The discussion of both matter and ghost sectors
can be carried out in parallel. The additional terms appearing in the
ghost part will contribute in the critical dimension to the 
cancellation of the Laplacian determinant coming from the ghost and coordinate
gluing. This point will be discussed later. The complete vertex
is given by the product of both pieces, the one corresponding to the matter
sector and that corresponding to the ghost, namely

$$
V (1, \cdots, N) = V_x (1, \cdots, N) V_\phi (1, \cdots, N).
$$

The general form of the Neumann coefficients $N_{n,m}^{r,s}$ have
been calculated in many works \cite{ASE}, for our purpose
it is convenient the representation given in \cite{JBA} since, for
any vertex, they can be written in a compact form in
a N-dimensional space spanned by the $N$ strings, as elements
of $N \times N$ matrix. For instance, one has for the terms which
do not involve the zero modes
$$
\begin{array}{l}
N_{even \, even}^{r,s} =  ({\bf I}_N  M_1^T - {\bf S_+}  M_2^T)^{-1} \,
({\bf I_N} M_2^T - {\bf S_+} M_-^T),\\[4ex]
N_{odd \, odd}^{r,s} =  {\bf S_+} -   {\bf S_-}^T
({\bf I_N} M_1 - {\bf S_+} M_2)^{-1} \, M_2 {\bf S_-},\\[4ex]
N_{odd \, even}^{r,s} = {\bf S_-}^T
({\bf I_N} M_1 - {\bf S_+} M_2)^{-1}  \, 
\end{array}
$$

\noindent
The remaining ones are of similar form and do not add anything new
to these expressions. ${\bf I_N}$ is the identity N-dimensional matrix, 
whereas the matrices ${\bf S_+}$ and ${\bf S_-}$ are

$$
\begin{array}{ll}
({\bf S_+})_{ij} = & \frac{1}{2} (1 + \delta_{ij}),\\[2ex]
({\bf S_-})_{ij} = & \frac{(-1)^{i+1}}{2} (1 - \delta_{ij}).\\
\end{array}
$$

The infinite dimensional matrices $M_{1,2}$ give the canonical
transformation from the half string degrees of freedom to
the conventional ones. Explicitly they have the form \cite{CHA}
\begin{equation}
\begin{array}{l}
(M_1)_{n,m} = \frac{2}{\pi} \left( \frac{2n}{2m - 1} \right)^{1/2}
\; \frac{(-)^{n + m}}{2 n - (2m - 1)}\, ,\\[2ex]
(M_2)_{n,m} = \frac{2}{\pi} \left( \frac{2n}{2m - 1} \right)^{1/2}
\; \frac{(-)^{n + m}}{2 n + 2m - 1}\, .
\end{array}
\label{matrices}
\end{equation}

One can see that, after diagonalization of the ${\bf S_\pm}$ matrices,
the problem of computing the vertex  itself, essentially
reduces to calculating the inverse of the matrix
$(M_1 - cos \frac{2 k \pi}{N} M_2)$ where $k = 1, ..., N$. Details of 
the whole process as well as the general form of this matrix and its
inverse have been given in ref. \cite{JBA}.

Now, one can use this result in the proof of the Gluing Theorem \cite{ALE}.
The theorem states  that the contraction of two string vertices (e.g.,
for $N$ and $N'$ strings), through the BPZ inner product (which is 
nothing other than the vertex overlapping two strings, $< V (12)|)$, 
gives the $(N + N' - 2)$-string 
vertex. More precisely, using the notation of vertices given in
equations (1) and (3), including ghost and matter sectors in the same
`bra', the result can be cast 

\begin{eqnarray}
< V (1, ... N-1, a)| & < V (N, ..., N + N' - 2, a')|V (a, a')> = 
\nonumber \\
& = < V (1, ..., N + N' - 2)| .
\label{vertex}
\end{eqnarray}

This  means that we have contracted the two vertices by joining
the strings $a$ and $a'$ in the first and the second vertex
respectively. Of course, because of the cyclic  properties of the
vertices, it is immaterial the position of these strings.

In terms of the Neumann representation of the vertex given in equations
(2) and (4),
the theorem shows the identification of the coefficients appearing
in $V (1, ..., N + N' - 2)$ with particular combinations of the ones
appearing in  $V (1, \cdots, N - 1,a)$ and $V (1, \cdots,  N' - 1, a')$.
In particular, in the proof of the theorem, and in the coordinate
sector, one has to show the identifications

\begin{eqnarray}
{\bf N}^{s_1, s_2} & = & N^{s_1,s_2} +  N^{s_1, a}  
 E N^{a', a'} E (1 - N^{a,a}  EN^{a',a'} E)^{-1}
N^{a,s_2} , 
\nonumber \\
{\bf N}^{s'_1, s'_2} & = & N^{s'_1,s'_2} +  N^{s'_1, a'}  
 E N^{a, a} E (1 - N^{a',a'}  EN^{a, a} E)^{-1}
N^{a', s'_2}, 
\nonumber \\
{\bf N}^{s_i, s'_j} & = & N^{s_i, a} -    
 E (1 - N^{a',a'}  EN^{a, a} E)^{-1}
N^{a',s'_j},
\label{neumann1}
\end{eqnarray}

\noindent
where   the indices $s_i (s'_i)$ run from 1 to $N -1$  ($N$ to $N' + N - 2$).
The bold-faced ${\bf N}$ represent the coefficients of the  resulting
glued vertex
according to the index. The matrix $E_{n, m} = (-)^{(n + 1)} \delta_{n,m}
(n, m \geq 1)$ comes from the BPZ inner product that, in the language 
of vertices, is $V (a, a')$. Also, after the contraction of the vertices with
the BPZ inner product \cite{ABE}, the determinant of the Laplacian appears.
This amounts to a factor of the form
\begin{equation}
{\bf D_x } = [det (1 - N^{a' a'} EN^{a, a} E)]^{-D/2} , \, \,
\label{determinant}
\end{equation}

\noindent
$D$ being the space time dimension, this factor will cancel once the 
contribution from the ghost sector is included.

Hence, the proof of the theorem relies on the calculation of the inverse
matrix $(1 - N^{a, a} EN^{a,'a'} E)^{-1}$.  Due to the symmetries of the 
diagonal Neumann coefficients there is no dependence on the indices $a$ 
and $a'$.
The general form of 
this inverse matrix can be obtained in full; the algebra
involved is rather tedious, and the result
 can be cast in terms
of the matrix $M_1^{-1} M_2 = \phi$. For instance, for 
an equal and even number of strings in each vertex, i.e. $N = N'$ $even$,
one can show that
\begin{eqnarray}
\begin{array}{ll}
& (1 - N^{a,a} EN^{a', a'} E)^{-1} = 
\nonumber \\
= & 2 \sqrt{1 - \phi^2} \, ( 1 + \sqrt{1 - \phi^2}) 
\frac{
\left( 1 + \sqrt{1 + \phi^2} \right)^{2N -2} - \phi^{2N-2}}
{\left( \left( 1 + \sqrt{1 - \phi^2} \right)^N
- \phi^N \right)^2} \, .
\end{array}
\label{generalresult}
\end{eqnarray}

This formula   is not at all illuminating, to illustrate
the essential points in the proof, we will work out in detail the
particular case of the gluing  of two 3-string vertices to
give the 4-string vertex. This vertex also appears explicitly in
the String Field Theory Action \cite{EWI} for open strings and, as it was
commented in the introduction, there is a possibility for
generating the non-polynomial string field theory action using
only this vertex. 

More precisely we are interested in the relation
$$
< V (1,2, a)| < V (a', 3,4)| V(a,a')> = < V (1,2,3,4)|,
$$
\noindent
which, in the momentum basis,  and restricting ourselves  to the
coordinates degrees of freedom is written as:
$$
\begin{array}{l}
\int \Pi_{i =1}^4  d k_i \, d k_a \, d k_{a'} \; 
\delta \left( k_a - k_1 - k_2 \right)
\, 
\delta \left( k_{a'} - k_3 - k_4 \right)
 \, 
\delta \left( k_a - k_{a'} \right)\\[2ex]
< k_1| < k_2| < k_a| V_x (1, 2, a) < k_{a'}| < k_3| < k_4| \, 
V_x (a', 3,4)
\, 
V_x (a, a') |k_{a'} > |k_{a'}> \\[2ex]
= 
\int \Pi^4_{i = 1} d k_i \, 
\delta \left( \sum_{i = 1}^4 k_i \right) \, 
< k_1|  < k_2 | < k_3| < k_4| V_x (1, 2, 3, 4).
\end{array}
$$

We focus our attention on the nonzero modes. In the coordinate sector,
when zero indices are involved, no difficulties arise and the proof
can be carried out along the same lines, the additional linear
term appearing in  the ghost vertex will be considered below. For the
$N = 3$ vertex, one has for the nonzero modes:

\begin{equation}
\begin{array}{l}
N^{r,r}_{even, \, even} = \frac{1}{2} (M_1^T + \frac{1}{2} M_2^T)^{-1} 
M_2^T  , \\[2ex]
N^{r, r \pm 1}_{even, \, even} = N^{r, r \pm 2}_{even, \, even} = - \frac{1}{2}
(M_1^T + \frac{1}{2} M_2^T)^{-1} (M_1^T + M_2^T) , \\ [2ex]
N^{r, r}_{odd, \, odd} = \frac{1}{2} (M_1 + \frac{1}{2} M_2)^{-1} 
M_2 , \\[2ex]
N^{r, r \pm 1}_{odd, \, odd} = N^{r, r \pm 2}_{odd, \, odd} = - \frac{1}{2}
(M_1 + \frac{1}{2} M_2)^{-1} (M_1 + M_2) , \\ [2ex]
N^{r,r}_{even, \, odd} = N^{r,r}_{odd,\, even} = 0, \\ [2ex]
N^{r, r \pm 1}_{even, \, odd} = N^{r,r \pm 1}_{odd, \, even} = - 
N^{r, r \pm 2}_{even, \, odd} = N^{r, r \pm2}_{odd, \, even} 
 =  ( M_1^T  +\frac{1}{2} M_2^T)^{-1} .
\end{array}
\label{neumann}
\end{equation}

In this case all the information is encoded in the inverse matrix given 
in (\ref{neumann1}), namely 
$$
M^{-1} = (1 - N^{a' a'} E N^{a,a} E)^{-1} ,
$$
\noindent
(notice that this matrix is effectively independent of the indices 
because of the symmetries of the Neumann coefficients).

To invert this matrix one needs only to make use of the properties
of the matrices $M_{1,2}$ dictated by the fact that they define
a canonical transformation in the string degrees of freedom \cite{CHA}. In
particular, they preserve the commutation relations of the string
creation and annihilation operators, hence one has
$$
\begin{array}{l}
M_1 M_1^T - M_2 M_2^T = I, \\
M_1 M_2^T - M_2 M_1^T = 0.
\end{array}
$$
A short calculation gives  the inverse matrix

\begin{eqnarray}
\left( M^{-1} \right)_{even,even} &=&
 \left( M_1+ \frac{1}{2} M_2 \right) \left( 1- M_1^{-1} M_2 \right)
\left( M_1^T+ \frac{1}{2} M_2^T \right),
\nonumber \\
\left(M^{-1}\right)_{even,odd} &=& \left(M^{-1}\right)_{odd,even}= 0,
\nonumber \\
\left(M^{-1}\right)_{odd,odd} &=&
\left( M_1^T+ \frac{1}{2} M_2^T \right) 
\left(1- M_2 M_1^{-1}\right) \left(M_1+ \frac{1}{2} M_2 \right).
\label{matrixm1}
\end{eqnarray}
\noindent

The final step is now to substitute this result in eq. (7) and
complete the form of the 4-vertex. In particular one gets
$$
\begin{array}{l}
{\bf N}^{r, r} = \frac{1}{2}  \left(
\begin{array}{cc}
M_2 M_1^{-1} & 0\\[3ex]
0 & - M_1^{-1} M_2 \end{array} \right)
\quad  , \quad 
{\bf N}^{r, r  + 1} = \frac{1}{2} \left( \begin{array}{cc}
- 1 & (M_1^{-1})^T\\[3ex]
- M_1^{-1} & 1 \end{array} \right) ,
\end{array}
$$

$$
{\bf N}^{r,r + 2} = \frac{1}{2} \left( \begin{array}{cc}
M_2 M_1^{-1} & 0 \\[2ex]
0 & - M_1^{-1} M_2 \end{array} \right) \quad , \quad
{\bf N}^{r, r + 1} = \frac{1}{2} \left( \begin{array}{cc}
- 1 & (M_1^{-1})^T \\[2ex]
- M_1^{-1} & 1 \end{array} \right) \, ,
$$

\noindent
to be compared with the result given in \cite{JBA} for the general vertex
(we have arranged rows and columns according to the parity of the 
indices as in equation (\ref{neumann})).

Other terms relating the remaining Neumann functions involve the same
sort of algebra and we will not insist on producing the calculation here.
It is important to remember that, in the process of contracting the oscillators
modes, the determinant term given in equation (\ref{determinant})  appears,
which will have to be combined with terms coming from the ghost sector.

Contraction of the oscillator modes in the ghost sector is performed
using the same techniques. In this case, however, because of the linear
factor $K^s_n$, there are additional contributions proportional to the
combinations of the zero modes or, equivalently, to the total central
charge $Q$.

In fact, apart from the vertex operator of the glued vertex, one obtains
the above mentioned term which, combined with the determinant of the Laplacian
from the coordinate sector (\ref{determinant}) amounts to a factor
\begin{equation}
{\bf D_x }{\bf D_\phi } e^{\frac{1}{2}Q^2 {\bf k}},
\label{globalfactor}
\end{equation}
\noindent
{\bf k} is given by a  combination of the matrices (\ref{matrices})
and the vector (\ref{vector}), whose explicit form is not needed for our
purpose, and the determinant of the Laplacian, the additional piece coming
from the contraction of the ghost operators using the BPZ inner product
now appears to the power $\frac{-1}{2}$
\begin{equation}
{\bf D_\phi } = [det (1 - N^{a' a'} EN^{a, a} E)]^{-1/2} \, \,
\label{determinant1}
\end{equation}

To proof that this factor (\ref{globalfactor}) is  equal to $1$, we proceed in 
the following way: first consider the state generated by the moments of the
energy-momentum operator acting on the momentum and ghost number eigenstate
$$
< a | = <k,-q-Q| exp \frac{1}{2} \left\{ a_n N^x_{n,m} a_m + \phi_n 
\tilde{N}_{n,m}^\phi \phi_m \right\},
$$
\noindent
(with an obvious notation for the Neumann coefficients)
Due to the fact that 
the central charge of this operator is zero in $D=26$ space-time
dimensions and $Q=-3$, we have, on the sphere,
the vanishing of all  moments of it. In the language of
vertices this means
$$
< a | < a' | V (a, a') |k_a > | k_{a'} > = 1 \, .
$$

Now, calculating explicitly the left hand side of the former relation,
one arrives to the identity
$$
1 = {\bf D_x}{\bf D_\phi} e^{\frac{1}{2}Q^2 {\bf k}},
$$
\noindent
where ${\bf k}$ is the same contribution as the one appearing in 
(\ref{globalfactor}), this derives from the fact that the vector
$\vec{K}^r$ is independent of the string under consideration, namely the value
of $r$, thus the $Q$ and $D$ dependence are the same in both equations.
The extra coefficient in front of the glued vertex equals one thus 
proving the validity of the theorem.

This result completes the proof of the theorem. Other vertices can be glued
following these lines, however the calculational details
 are more involved. This is easily seen in the form of equation 
(\ref{generalresult}).

Now, we want to argue that the Gluing theorem can be applied to calculate
actual string amplitudes. It has been shown in \cite{CHA} that string
amplitudes themselves can be represented as the sum of contact string
interactions in all possible reparametrization of the strings. Also
in \cite{NONPOL} it was shown that the terms of the non-polynomial closed
string action can be obtained as a sum of string contact interactions, in
a region of the reparametrization modular space appropriate to reproduce
the so called restricted polyhedra. With this two ideas in mid, on can
imagine the gluing of vertices in order to reproduce string amplitudes.
For instance, starting with the three strings vertex and applying the 
reparametrization operator $(\Omega)$  defined in \cite{CHA} to each
vertex, one could envisage the construction of string amplitudes
starting from a generalization of the three string vertex. More precisely
one has

$$
\begin{array}{ll}
< V (1,2, a) | \left( \Pi_{i=1,2,a} 
\, \Omega_i \right) & <
V (a',3,4) | \left( \Pi_{i = a',3,4}  \, \Omega_i \right) | V (a, a') >
= \\[2ex]
& < V (1,2,3,4) | \left( \Pi_{i = 1, ...,4} \, \Omega_i \right)
\end{array}
$$

\noindent
where we have, with the BPZ inner product, glued two vertices taking all
possible reparametrization of the strings to give a four string vertex
which, in reference \cite{CHA} was shown to reproduce the tree level 
string amplitudes for both closed and open strings.

Work in this directions is in progress and we hope to report on it in
the near future. This result will open the way to simplifying the
construction of amplitudes in the case of closed strings where in the
non-polynomial theory of \cite{MKA} requires the addition of terms in every
order of perturbation theory.

To summarize, in this note we have given an algebraic proof of the
Gluing Theorem of string vertices applied to the String Field Theory due to
Witten \cite{EWI}. We have used the representation of Neumann functions given
by the half-string formulation of the theory and the problem is
reduced to the calculation of inverse matrices (of infinite dimension).
The introduction of the reparametrization ghosts allows us to get rid
of the determinant of the Laplacian coming from the operator contraction
in the process of gluing. Finally we outlook the possibility that
the theorem, when applied to a suitable generalization of the string
vertices, can generate the open and closed string amplitudes. This
result will greatly simplify the construction of amplitudes in the
context of the non-polynomial string field theory.

One of us (JB) is supported in part by grants CYCIT96-1718, PB97-1261 
and GV98-1-80.  He would also like to thank the Rutherford Appleton
Laboratory for hospitality.

\end{document}